\documentclass[12pt,draftclsnofoot,journal,onecolumn]{IEEEtran}
\usepackage{graphicx}
\usepackage{amsmath}
\usepackage{array}
\newcommand{\PreserveBackslash}[1]{\let\temp=\\#1\let\\=\temp}
\newcolumntype{C}[1]{>{\PreserveBackslash\centering}p{#1}}
\newcolumntype{R}[1]{>{\PreserveBackslash\raggedleft}p{#1}}
\newcolumntype{L}[1]{>{\PreserveBackslash\raggedright}p{#1}}
\usepackage[usenames]{color}
\usepackage{colortbl,booktabs}
\usepackage{tabularx}
\usepackage{multicol}
\usepackage{booktabs}
\usepackage[subfigure]{tocloft}
\usepackage{subfigure}
\usepackage{bm}
\usepackage{cite}
\usepackage{balance}

\usepackage{algorithm}

\usepackage{threeparttable}
\usepackage{amsmath}
\usepackage{dcolumn}
\usepackage{multirow}
\usepackage{booktabs}
\usepackage{amsfonts}
\usepackage{framed}
\newcolumntype{d}[1]{D{.}{.}{#1}}
\begin{document}

\bibliographystyle{IEEEtran} 
\title{Reconfigurable Intelligent Surface Based Hybrid Precoding for THz Communications}
\author{{{Yu Lu, {Mo Hao}, and {Richard MAcKenzie}}}

		
}
\maketitle
\begin{abstract}

Benefiting from the growth of the bandwidth, Terahertz (THz) communication can support the new application with explosive requirements of the ultra-high-speed rates for future 6G wireless systems. In order to compensate for the path loss of high frequency, massive multiple-input multiple-output (MIMO) can be utilized for high array gains by beamforming. However, since a large number of analog phase shifters should be used to realize the analog beamforming, the existing THz communication with massive MIMO has very high energy consumption. To solve this problem, a reconfigurable intelligent surface (RIS)-based hybrid precoding architecture for THz communication is developed in this paper, where the energy-hungry phased array is replaced by the energy-efficient RIS to realize the analog beamforming of the hybrid precoding. Then, based on the proposed RIS-based architecture, a sum-rate maximization problem for hybrid precoding is investigated. Since the phase shifts implemented by RIS in practice are often discrete, this sum-rate maximization problem with a non-convex constraint is challenging. Next, the sum-rate maximization problem is reformulated as a parallel deep neural network (DNN)-based classification problem, which can be solved by the proposed low-complexity deep learning-based multiple discrete classification (DL-MDC) hybrid precoding scheme. Finally, we provide numerous simulation results to show that the proposed DL-MDC scheme works well both in the theoretical Saleh-Valenzuela channel model and practical 3GPP channel model. Compared with existing iterative search algorithms, the can proposed DL-MDC scheme reduces the runtime significantly with a negligible performance loss.

\end{abstract}


\section{Introduction}\label{S1}

\IEEEPARstart 6G is expected to achieve about a 100-fold increase in data rate than 5G~\cite{TS19,6G1,6G_letaief}. Due to the tremendous growth of the bandwidth, Terahertz (THz) communication has been regarded as a promising technology for future 6G wireless systems to support the high data rate application. Compared with the typical bandwidth of several GHz in the millimeter-wave (mmWave) band (30-100 GHz) for 5G~\cite{mmWave2014}, the THz band (0.1-10 THz) is able to provide tens of GHz bandwidth~\cite{6G1}. 
One of the major obstacles for THz communication is the severe propagation attenuation at high frequencies~\cite{TS19}. Similar to mmWave communication, in order to alleviate the severe signal attenuation problem, THz communication can also use hybrid precoding based on massive multiple-input multiple-output (MIMO) to generate directional beams with high array gains~\cite{6G_letaief}.

However, the direct application of hybrid precoding in THz communication will result in the very high energy consumption mainly due to the following reason. It is well known that the phased array based hybrid precoding requires a large number of analog phase shifters to realize the analog beamforming in mmWave communication, and this number in THz communication will be significantly increased, since much more antennas should be used to compensate for the more severe attenuation of THz signals than mmWave signals.  For instance, 128 or 256 antennas are usually considered in mmWave systems, while 1024 or 2048 antennas have been considered in THz systems~\cite{chen19}.

Fortunately, the recently proposed reconfigurable intelligent surface (RIS) is regarded as a revolutionary technique to provide an energy-efficient alternative to the traditional energy-hungry phased array~\cite{zhangs19}. Specifically, RIS is an artificial metasurface consisting of a large number of low-cost and passive elements. Each element on RIS is able to reflect the incident signal while adjusting its phase shift individually~\cite{zhangW19,Hu19,ES_1bit}. The elements on RIS are nearly passive without the need of dedicated radio frequency (RF) chain, so they can realize phase shifts in an energy-efficient way. By individually manipulating the phase shifts of elements on RIS in a preferred way, RIS is able to generate dynamic directional beams to different users in real time according to the changing wireless channels and mobile users. Note that RIS is also different from the existing lens antenna array, which can only provide fixed beams with fixed phase shifts~\cite{WangBC19}. Thus, RIS is a potential technique to address the high energy consumption problem for THz communication.

\subsection{Prior works}\label{S1.1}
The main applications of RIS in wireless communication systems can be divided into two categories: a reflective relay, and an antenna array.

First, RIS can be used as a reflective relay to reconstruct the channel link blocked by buildings or other objects, which enhances the wireless link robustness and optimizes the link outage probability~\cite{transmitter2}, particularly for mmWave and THz communications. Moreover, RIS also has some other advantages, such as increasing the coverage area, enriching the channel with larger number of multi-paths, reducing the transmit power, etc.~\cite{Yuen2}. Some recent works have investigated the performance of the wireless system with RIS acting as a relay~\cite{transmitter2,Wu'18,Chen2020}. These works mainly concentrate on the optimal RIS beamforming design to enhance the performance of wireless communication. Specifically, the authors in~\cite{Wu'18} proposed an alternating optimization for a single RIS to maximize the signal-to-noise ratio (SNR) of a single user as well as multiple users, where the ideal continuous phase shifts of the RIS are considered. Furthermore, the optimal RIS beamforming design was investigated for multiple RISs, which is also based on the continuous phase shifts~\cite{Hanzo19}.

However, the continuous phase shifts on RIS are challenging to realize in practice due to the hardware constraint of RIS, e.g., the most recently reported RIS prototype at mmWave frequency 28 GHz can realize up to 2-bit discrete phase shifts~\cite{Dai19}. Fortunately, it has been proved that the gap of the sum-rate performance between the discrete phase shift (even for the 1-bit phase shifter) and the continuous phase shifter is negligible, and it keeps constant when the number of RIS elements approaches infinity~\cite{17zhang19}. Thus, we are likely to use RIS with discrete phase shifts for practical wireless communications. The passive RIS beamforming optimization problem with discrete phase shifts is non-convex. The exhaustive search can be used to find out the optimal solution to this non-convex problem, but the complexity is very high.
The main way of reducing the complexity is to utilize the iterative search algorithms. For example, the authors have proposed the iterative search method for the RIS beamforming optimization to reduce the transmit power in~\cite{17zhang19} and increase the sum-rate in~\cite{Al-Dhahir19}, respectively. However, this kind of iterative search schemes still suffers from the high complexity when the RIS dimension is large.

The second application of RIS is an antenna array, particularly equipped at the base station (BS). Compared with the typical multi-antenna transceiver, the RIS-aided transceiver has much lower energy consumption and hardware cost~\cite{Yuen3}. There are only very limited initial works for this kind of application for RIS~\cite{transmitter1,transmitter3}. Specifically, in~\cite{transmitter1}, the digital baseband signals are directly mapped to the control signals on RIS to change the reflection coefficients of RIS elements, so the RF modulation of the reflected electromagnetic waves can be achieved. It has been shown that a low-cost multi-stream transmitter is able to be implemented by a single-antenna feeder and an RIS. Furthermore, the architecture with a single-antenna feeder was proposed in~\cite{transmitter3}, where the RIS is an integrated part of the transmitter at the BS, and it is claimed that the antenna feeder can be directly connected with the RIS transmitting modulated signals instead of reflecting them. However, these works use RIS to modulate the RF signals by simultaneously (not individually) controlling all RIS elements via the same control signals, so it is not easy to realize analog beamforming with high array gains.

\subsection{Our contributions }\label{S1.2}
In this paper, we first propose the RIS-based hybrid precoding architecture for THz communication with low energy consumption. 
The key idea of this architecture is to replace the energy-hungry phased array in the conventional hybrid precoding architecture by the energy-efficient RIS to realize the analog beamforming. Then, for the proposed RIS-based hybrid precoding architecture, we formulate a sum-rate maximization problem, which is proved to be a classification problem.
We further propose a deep learning-based multiple discrete classification (DL-MDC) hybrid precoding scheme to solve this classification problem with reduced complexity. The specific contributions can be summarized as follows.
\begin{itemize}

 \item Unlike most of the existing works that employ RIS at the transmitter to modulate RF signals without array gains, we utilize the energy-efficient RIS to replace the phased array to realize analog beamforming in THz communication systems in the proposed RIS-based hybrid precoding architecture. By individually controlling a great number of energy-efficient and low-cost passive elements instead of expensive phase shifters, the RIS in the RIS-based hybrid precoding architecture is able to reflect signals in directional beams to different users. 
 
 \item For the proposed RIS-based hybrid precoding architecture, considering discrete phase shifts due to the practical hardware implementation of RIS, a sum-rate maximization problem is formulated in RIS-based THz communication. This non-convex constraint makes the sum-rate maximization problem non-convex, so the optimization of the tunable discrete phase shifts of RIS to realize the optimal analog beamforming is challenging. Instead of using iterative search methods with high complexity to achieve the solution to this non-convex problem, the mapping relationship is analyzed between the channel matrix and the analog beamforming matrix at first, based on which the non-convex sum-rate maximization problem is reformulated as multiple discrete classification problems.
 
 \item We propose a DL-MDC algorithm to solve the reformulated sum-rate maximization problem by leveraging the excellent performance of deep neural network (DNN) for handling the classification problems. Specifically, the 1-bit phase shifts of RIS are considered in the proposed RIS-based hybrid precoding architecture, which can be easily implemented via diodes with very low power and cost. 
 Instead of using one DNN directly, we develop a parallel DNN system to decide every non-zero element in the analog beamforming matrix, where each DNN of the parallel DNN system is constructed with only one output. The output of DNN is the posterior probability of phase shift being 0. Then, a reasonable threshold is set for the output to obtain the classification outcome. It is worth pointing out that, since the channel direction information is more important to generate directional beams than the channel gain information for analog beamforming, we preprocess the data samples based on the physical property of the hybrid precoding to take full advantage of DNN. 
 
  \item We provide the simulation results of the proposed DL-MDC hybrid precoding scheme both in the theoretical Saleh-Valenzuela channel model~\cite{Li18} and the practical 3GPP channel model~\cite{3GPP}. It is shown that the proposed scheme is able to predict the classification results precisely. Meanwhile, the near-optimal sum-rate performance can be achieved with low complexity.

\end{itemize}

\subsection{Organization and notation}\label{S1.3}
{\it Organization}: The rest of the paper is organized as follows. Section II describes the proposed RIS-based hybrid precoding architecture for THz communication at first, and then the sum-rate maximization problem is formulated under the hardware constraint of the proposed architecture. In Section III, a low-complexity DL-MDC hybrid precoding scheme for THz communication is proposed to address the non-convex sum-rate maximization problem. Simulation results are provided in Section IV. Finally, some conclusions are drawn in Section V.

{\it Notation}: Lower-case and upper-case boldface letters ${\bf{a}}$ and ${\bf{A}}$ denote a vector and a matrix, respectively; ${{{\bf{A}}^H}}$, ${{{\bf{A}}^T}}$, and ${{{\bf{A}}^{ - 1}}}$ denote the conjugate transpose, transpose, and inversion of the matrix ${\bf{A}}$ respectively; ${{\left\|  \bf{A}  \right\|_F}}$  denotes the Frobenius norm of the matrix ${\bf{A}}$; ${\rm{Diag}\left( \bf{A}\right) }$ denotes a column vector composed of
the diagonal elements  of ${\bf{A}}$; ${\rm{diag}\left( \bf{a}\right) }$ denotes the diagonal matrix whose diagonal elements consist of the elements in the vector ${\bf{a}}$; ${{\left\|  \bf{a}  \right\|_2}}$ denotes the ${{l_2}}$-norm of the vector ${\bf{a}}$; ${\left|  a  \right|}$ denotes the amplitude of the scalar ${a}$; We use ${\mathcal{C}\mathcal{N}(m,R)}$ to denote the complex Gaussian distribution with mean ${m}$ and covariance ${R}$, and ${\mathcal{U}(a, b)}$ to denote the uniform distribution in the range (a, b); ${\bf{\cal A}}$ denotes a set and ${\rm{card}({\bf{\cal A}})}$ denotes the number of elements in the set ${{\bf{\cal A}}}$; Finally, ${{{\mathbf{I}}_{K}}}$ is the ${K \times K}$ identity matrix, and ${\otimes }$ denotes the kronecker product.

\section{Proposed RIS-Based Hybrid Precoding Architecture}\label{S2}
In this section, we first introduce the system model of the proposed RIS-based hybrid precoding architecture for THz communication, where the analog beamforming of hybrid precoding will be realized by energy-efficient RIS instead of the energy-hungry phased array.
Then, based on the proposed architecture, we formulate the optimization problem of hybrid precoding to maximize the sum-rate of all users with a non-convex constraint due to the discrete phase shifts in practical  hardware implementation of RIS.

\subsection{System model}\label{S2.1}

In this paper, we consider a typical single-cell downlink THz massive MIMO system, where the BS employs ${M}$ RF chains to serve ${K}$ single-antenna users. ${M = K}$ is assumed to fully exploit the multiplexing gain~\cite{Alkhateeb1}.

\begin{figure}[tp]
	\begin{center}
		\vspace*{0mm}\includegraphics[scale=0.5]{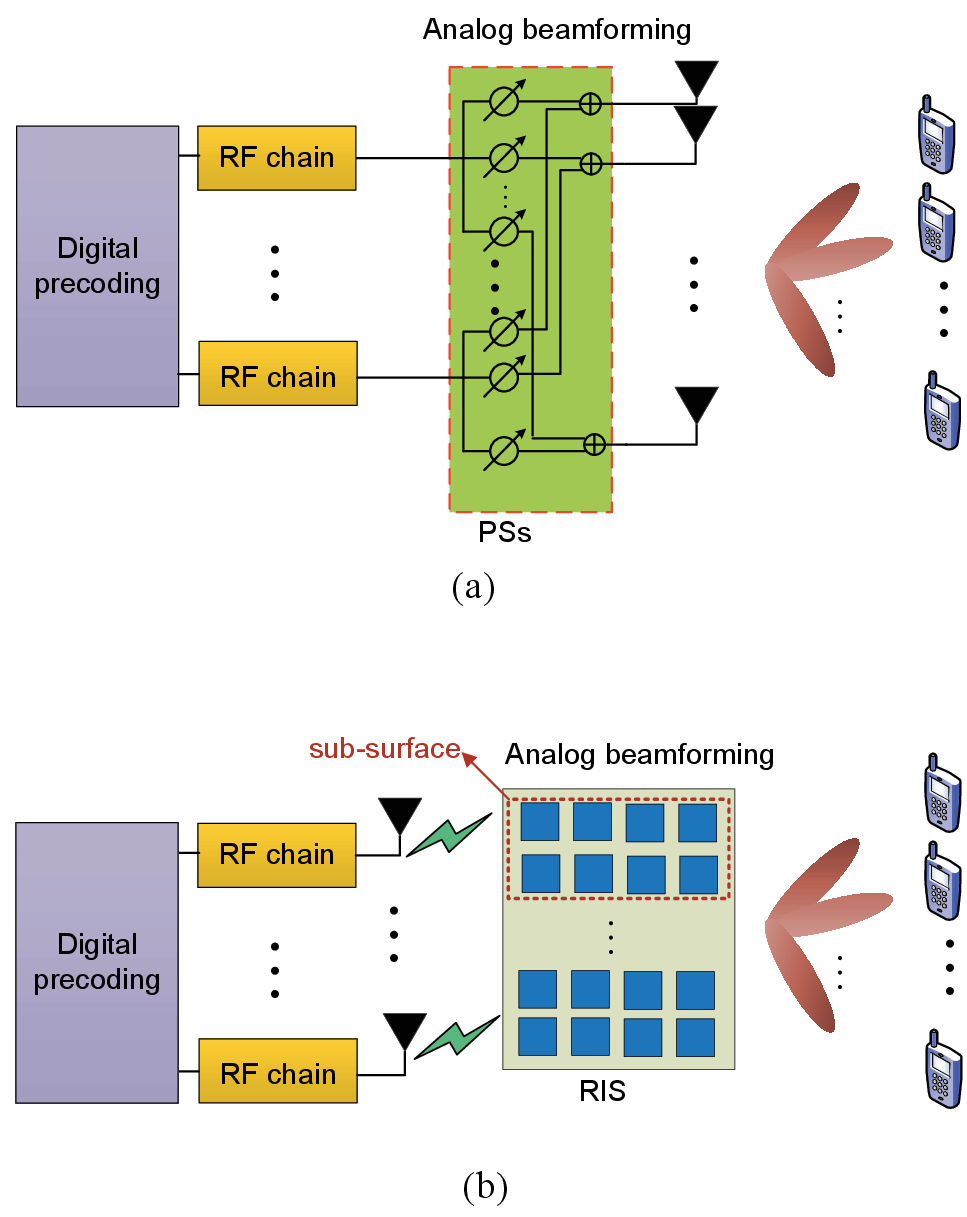}
	\end{center}\vspace*{-3mm}
	\vspace*{-3mm}\caption{Comparison of two hybrid precoding architectures: (a) Conventional phased array based hybrid precoding; (b) Proposed RIS-based hybrid precoding.}
	\vspace*{3mm}\label{V2_1}
\end{figure}

In the conventional phased array based hybrid precoding architecture illustrated in Fig. 1 (a), the RF chains are connected to the phased array for analog beamforming, and the number of required phase shifters is proportional to the number of RF chains as well as the number of antennas for the widely considered fully-connected scheme~\cite{Ayach2014hybrid}.

By virtue of the low-rank characteristics of mmWave/THz channels~\cite{Rappaport2013mmwave}, the phased array based hybrid precoding is able to achieve the near-optimal performance. However, in mmWave systems, the number of antennas at BS is usually 128 or 256, while in THz systems this number is much larger, e.g., 1024 or 2048 in~\cite{chen19}. Thus, due to numerous energy-hungry phase shifters in THz systems, the conventional phased array based hybrid precoding architecture  will result in very high energy consumption. 

As shown in Fig. 1 (b), the proposed RIS-based hybrid precoding architecture for THz communication utilizes the energy-efficient RIS rather than the energy-hungry phased array for analog beamforming, where the RF signals are reflected by the RIS. Particularly, RIS is an artificial metasurface consisting of a large number of low-cost passive elements. These elements have the ability to generate dynamic directional beams with high array gains by individually manipulating their phase shifts~\cite{ES_yang}. 
Thanks to the passive RIS elements, compared to the conventional phased array based hybrid precoding architecture, the proposed RIS-based hybrid precoding architecture is capable to realize the analog beamforming in a more energy-efficient way.

Specifically, the RIS employs ${N}$ passive elements, and each RF chain is only associated with a sub-surface consisting of a non-overlappping part of the passive elements in the proposed RIS-based hybrid precoding architecture, as shown in Fig. 1 (b). 
Then, we assume that there are $M$ sub-surfaces, and the signal from each RF chain is reflected by a unique sub-surface consisting of $N_s=N/M$ passive elements ($N/M$ is assumed to be an integer for simplicity). Note that by carefully designing the size of the sub-surfaces and the distance between the RF chains and the RIS, the interference among different sub-surfaces can be ignored~\cite{Guan19}, which means that each RF chain's signal can be only reflected by a sub-surface with $N_s$ passive elements.

The signal ${y_k}$ received at the ${k}$th user is expressed as
\begin{equation}\label{eq1}
{{y_k} = \sum\limits_{m = 1}^M {{{\bf{h}}_{k,m}^H}{{\bf{\Phi }}_{m}}{{\bf{g}}_m}{{\bf{f}}_{m}^{T}}{\bf{s}}\ {\rm{ + }}\ n}},
\end{equation}
where ${{\bf{s}} \in {\mathbb{C}}^{K \times 1}}$ is the transmitted signal vector satisfying ${{\mathbb{E}}\left[{{\bf{s}}{{\bf{s}}^{H}}}\right] = {\bf{I}}_{K}}$ for all ${K}$ users, and ${{{\bf{f}}_{m}}} \in {\mathbb{C}}^{K \times 1}$  is the ${m}$th column of ${{\bf{F}}^T}$, where ${{\bf{F}}=[{{\bf{f}}_{1}}, {{\bf{f}}_{2}}, \cdots ,{{\bf{f}}_{{M}}}]^T \in {\mathbb{C}}^{M \times K}}$ is the digital precoding matrix, and ${{\bf{g}}_{m}} \in \mathbb{C}^{{N_s}\times 1}$  is the beamforming channel vector between the $m$th antenna feeder and the $m$th sub-surface; The analog beamforming matrix $ \mathbf{\Phi} _{m} \buildrel \Delta \over = \operatorname{diag}\left[\phi_{m,1}, \phi_{m,2}, \cdots, \phi_{m,N_{s}}\right] \in {\mathbb{C}}^{{N_s}\times {N_s}}$ is a diagonal matrix, where $ \phi_{m, n_s}=e^{j \theta_{m, n_s}} $ is the phase shift of the $ n_s $th element on the $m$th sub-surface; ${{\bf{h}}_{k,m}} \in \mathbb{C}^{{N_s}\times 1}$ is the channel vector between the $m$th sub-surface and the $k$th user; $n$ is the additive white Gaussian noise (AWGN) following the distribution $\mathcal{C}\mathcal{N}\left( 0,\sigma^{2} \right)$.

It is worth pointing out that two kinds of channel models in THz massive MIMO systems are considered: the theoretical Saleh-Valenzuela channel model and the practical 3GPP channel model, which will be discussed in detail in Subection II-B.

Considering the received signals ${\left\lbrace y_k\right\rbrace }_{k=1}^K$ for all $K$ users, the received signal ${\bf{y}}=\left[ y_1,y_2,\cdots,y_K\right]^T  $ can be rewritten as
\begin{equation}\label{eq3}
{{\bf{y}} = {{\bf{H}}{{\bf{\Phi }}{\bf{G}}{{\bf{F}}}{\bf{s}}}} + {\bf{n}}},
\end{equation}
where
\begin{equation}\label{eqx2}
{{\bf{G}} = {\left( {\begin{array}{*{20}{c}}
			{\bf{g}}_{1} & \ldots &0\\
			\vdots & \ddots & \vdots \\
			0& \cdots &{\bf{g}}_{M}
			\end{array}} \right)_{{N} \times {M}}}},
\end{equation}
${{\bf{H = }}{\left[{{{\bf{h}}_1},{{{\bf{h}}_2}},\cdots,{{{\bf{h}}_K}}}\right]^H} \in \mathbb{C}^{K \times N}}$ is the channel matrix with ${{\bf{h}}_k = [{\bf{h}}_{k,1}^T,{\bf{h}}_{k,2}^T,\cdots,{\bf{h}}_{k,M}^T]^T}$, ${{\bf{\Phi}}} \in \mathbb{C}^{N \times N}$ is the analog beamforming matrix expressed as
\begin{equation}\label{eq8}
{{\bf{\Phi }} = {\left( {\begin{array}{*{20}{c}}
			{\bf{\Phi }}_{1} & \ldots &0\\
			\vdots & \ddots & \vdots \\
			0& \cdots &{\bf{\Phi }}_{M}
			\end{array}} \right)_{{N} \times {N}}}}.
\end{equation}
To maximize the spectral efficiency of the RIS-based hybrid precoding architecture, the digital precoding matrix ${{\bf{F}}}$ and the analog beamforming matrix ${{\bf{\Phi}}}$ in (\ref{eq3}) should be jointly designed, which will be discussed in Subsection II-C.

\subsection{Channel models}
In this part, the theoretical Saleh-Valenzuela channel model and the practical 3GPP channel model for THz communication will be introduced.
\subsubsection{Saleh-Valenzuela channel model}
We first consider the typical theoretical Saleh-Valenzuela channel model in THz massive MIMO systems~\cite{Li18}, where the channel vector ${{\bf{h}}_{k,m}}$ in (\ref{eq1}) be presented as
\begin{equation}\label{eq4}
{{{\bf{h}}_{k,m}} = \sqrt {\frac{N}{{{L_{k,m}}}}} \sum\limits_{l = 0}^{{L_{k,m}}} {\alpha _{k,m}^{(l)}{\bf{a}}(\varphi _{k,m}^{(l)},\theta _{k,m}^{(l)})}},
\end{equation}
where ${L_{k,m}}$ is the number of multipath between the ${m}$th sub-surface and the ${k}$th user, ${{\bf{\alpha}} _{k,m}^{(l)}}$ denotes the complex gain of the ${l}$th path ${(1 \le l \le {L_{k,m}})}$, ${\varphi _{k,m}^{(l)}}$ and ${\theta _{k,m}^{(l)}}$ represent the azimuth and elevation angle of departure (AoD) of the ${l}$th path, ${{\bf{a}}(\varphi _{k,m}^{(l)},\theta _{k,m}^{(l)})} \in \mathbb{C}^{N_s \times 1}$ is the array response vector. When we consider the widely used uniform planar array (UPA) with ${N_{s1}}$ antenna elements in the horizontal direction and ${N_{s2}}$ antenna elements in the vertical direction (${N_s = {N_{s1}} \times {N_{s2}}}$), the array response vector can be presented as
\begin{equation}\label{eq5}
{{\bf{a}}(\varphi ,\theta )\;{\rm{ = }}\;{{\bf{a}}_{{\rm{az}}}}\left( \varphi  \right) \otimes {{\bf{a}}_{{\rm{el}}}}\left( \theta  \right)},
\end{equation}
where
\begin{equation}\label{eq6}
{{{{\bf{a}}_{{\rm{az}}}}\left( \varphi  \right){\rm{ = }}\frac{{\rm{1}}}{{\sqrt {{N_{s1}}} }}{[{e^{j2\pi i({d_1}/\lambda ){\rm{sin}}(\varphi )}}]^T}},\ {i \in {\cal I}\left( {{N_{s1}}} \right)}},
\end{equation}
\begin{equation}\label{eq7}
{{{{\bf{a}}_{{\rm{el}}}}\left( \theta  \right){\rm{ = }}\frac{{\rm{1}}}{{\sqrt {{N_{s2}}} }}{[{e^{j2\pi j({d_2}/\lambda ){\rm{sin}}(\theta )}}]^T},\ {j} \in {\cal I}\left( {{N_{s2}}} \right)}},
\end{equation}
where ${\lambda }$ is the signal wavelength, ${d_{\rm{1}}}$ (${d_{\rm{2}}}$) represents the element spacing in the horizontal (vertical) direction, and ${{\cal I}\left( n \right){\rm{ = \{ 0,1,}}\cdots{{,n - 1\} }}}$.

\subsubsection{3GPP channel model}
We also consider the practical 3GPP channel model in this paper. Specifically, the channel vector $ {{\bf{h}}_{k,m}} = \left[{h}_{k, n}(\tau, t)\right]^{T}  $, $n \in \left\lbrace (m-1)*N_s+1,\cdots,m*N_s\right\rbrace  $,  where $ {h}_{k, n}(\tau, t) $ is the channel between the $ n $th elements on RIS and the $ k $th user, which can be presented as:
\begin{equation}
{h}_{k, n}(\tau, t)=\sqrt{\frac{K_{R}}{K_{R}+1}} {h}_{k, n,1}^{\operatorname{\bm{LOS}}}(t) \delta\left(\tau-\tau_{1}\right)+\sqrt{\frac{1}{K_{R}+1}} {h}_{k, n}^{\mathrm{\bm{NLOS}}}(\tau, t),
\end{equation}
where $ K_{R} $ is the Ricean K-factor, $ \delta(\cdot) $ is the Dirac's delta function, $ {h}_{k, n,1}^{\operatorname{\bm{LOS}}} $ and $ {h}_{k, n}^{\mathrm{\bm{NLOS}}} $ denote the line of sight (LOS) channel component and non-line of sight (NLOS) channel component. The LOS channel component $ {h}_{k, n,1}^{\operatorname{\bm{LOS}}} $ is expressed as (\ref{eq20}) and the parameters are shown in~\cite{3GPP}.

\begin{equation}\label{eq20}
\begin{split}
{h}_{k, n,1}^{\operatorname{\bm{LOS}}}(t)=\left[\begin{array}{c}F_{{\rm{rx}}, k, \theta}\left(\theta_{\rm{LOS, ZOA}}, \phi_{\rm{L O S, A O A}}\right) \\ F_{r x, k, \phi}\left(\theta_{\rm{L O S, Z O A}}, \phi_{\rm{L O S, A O A}}\right)\end{array}\right]^{T}\left[\begin{array}{cc}1 & 0 \\ 0 & -1\end{array}\right]\left[\begin{array}{c}F_{{\rm{tx}}, n, \theta}\left(\theta_{\rm{L O S, Z O D}}, \phi_{\rm{L O S, A O D}}\right) \\ F_{{\rm{t x}}, n, \phi}\left(\theta_{\rm{L O S, Z O D}}, \phi_{\rm{L O S, A O D}}\right)\end{array}\right] \\ \cdot \exp \left(-j 2 \pi \frac{d_{3 D}}{\lambda_{0}}\right) \exp \left(j 2 \pi \frac{\hat{r}_{\rm{rx}, L O S}^{T} \cdot \bar{d}_{{\rm{r x}}, k}}{\lambda_{0}}\right) \exp \left(j 2 \pi \frac{\hat{r}_{\rm{tx}, L O S}^{T} \cdot \bar{d}_{{\rm{t x}}, n}}{\lambda_{0}}\right) \exp \left(j 2 \pi \frac{\hat{r}_{\rm{r x}, L O S}^{T}\cdot \bar{v}}{\lambda_{0}} t\right).
\end{split} 
\end{equation}

The NLOS channel component $ h_{k, n}^{\mathrm{\bm{NLOS}}} $ can be presented as:
\begin{equation}
h_{k, n}^{\mathrm{\bm{NLOS}}}(\tau, t)=\sum_{p=1}^{2} \sum_{i=1}^{3} \sum_{j \in R_{i}} h_{k, n, p, j}^{\mathrm{\bm{NLOS}}}(t) \delta\left(\tau-\tau_{p, i}\right)+\sum_{p=3}^{P} h_{k, n, p}^{\mathrm{\bm{NLOS}}}(t) \delta\left(\tau-\tau_{p}\right),
\end{equation}
where 2 multiple clusters channel $ h_{k, n, p, j}^{\mathrm{\bm{NLOS}}}(t) ~(p=1,2)$ and $ P-2 $ single cluster channel $ h_{k, n, p}^{\mathrm{\bm{NLOS}}}(t) ~(p=3,4,\cdots,P)$ are considered in this paper, which are shown as (12) and (13). The parameters of the NLOS channel component are presented in ~\cite{3GPP}.

\begin{equation}
\begin{array}{l}h_{k, n, p, j}^{\mathrm{\bm{NLOS}}}(t)=\sqrt{\frac{P_{p}}{J}}\left[\begin{array}{c}F_{{\rm{r x}}, k, \theta}\left(\theta_{p, j, \rm{Z O A}}, \phi_{p, j, \rm{A O A}}\right) \\ F_{{\rm{r x}}, k, \phi}\left(\theta_{p, j, \rm{Z O A}}, \phi_{p, j, \rm{A O A}}\right)\end{array}\right]^{T} \left[\begin{array}{c}\exp \left(j \Phi_{p, j}^{\theta \theta}\right) \\ \sqrt{\kappa_{p, j}^{-1}} \exp \left(j \Phi_{p, j}^{\phi \theta}\right) \\ \sqrt{\kappa_{p, j}^{-1}} \exp \left(j \Phi_{p, j}^{\theta \phi}\right) \\ \exp \left(j \Phi_{p, j}^{\phi \phi}\right)\end{array}\right] \\ {\left[\begin{array}{c}F_{{\rm{t x}}, n, \theta}\left(\theta_{p, j, \rm{Z O D}}, \phi_{p, j, \rm{A O D}}\right) \\ F_{{\rm{t x}}, n, \phi}\left(\theta_{p, j, \rm{Z O D}}, \phi_{p, j, \rm{A O D}}\right)\end{array}\right] \exp \left(j 2 \pi \frac{\hat{r}_{{\rm{r x}}, p, j}^{T} \cdot \bar{d}_{{\rm{r x}}, k}}{\lambda_{0}}\right) \exp \left(j 2 \pi \frac{\hat{r}_{{\rm{t x}}, p, j}^{T}\cdot \bar{d}_{{\rm{t x}}, n}}{\lambda_{0}}\right) \exp \left(j 2 \pi \frac{\hat{r}_{{\rm{r x}}, p, j}^{T} \cdot\bar{v}}{\lambda_{0}} t\right)}.\end{array}
\end{equation}

\begin{equation}
\begin{array}{l}h_{k, n, p}^{\mathrm{\bm{NLOS}}}(t)=\sqrt{\frac{P_{p}}{J}} \sum_{j=1}^{J}\left[\begin{array}{c}F_{{\rm{r x}}, k, \theta}\left(\theta_{p, j, {\rm{Z O A}}}, \phi_{p, j, \rm{A O D}}\right) \\ F_{{\rm{r x}}, k, \phi}\left(\theta_{p, j, \rm{Z O A}}, \phi_{p, j, \rm{A O A}}\right)\end{array}\right]^{T}\left[\begin{array}{c}\exp \left(j \Phi_{p, j}^{\theta \theta}\right) \\ \sqrt{\kappa_{p, j}^{-1}} \exp \left(j \Phi_{p, j}^{\phi \theta}\right) \\ \sqrt{\kappa_{p, j}^{-1}} \exp \left(j \Phi_{p, j}^{\theta \phi}\right) \\ \exp \left(j \Phi_{p, j}^{\phi \phi}\right)\end{array}\right] \\ {\left[\begin{array}{c}F_{{\rm{t x}}, n, \theta}\left(\theta_{p, j, \rm{Z O D}}, \phi_{p, j, \rm{A O D}}\right) \\ F_{{\rm{t x}}, n, \phi}\left(\theta_{p, j, \rm{Z O D}}, \phi_{p, j, \rm{A O D}}\right)\end{array}\right] \exp \left(\frac{j 2 \pi\left(\hat{r}_{{\rm{r x}}, p, j}^{T}\cdot \bar{d}_{{\rm{r x}}, k}\right)}{\lambda_{0}}\right) \exp \left(\frac{j 2 \pi\left(\hat{r}_{{\rm{r x}}, p, j}^{T}\cdot \bar{d}_{{\rm{r x}}, n}\right)}{\lambda_{0}}\right) \exp \left(j 2 \pi \frac{\hat{r}_{{\rm{r x}}, p, j} \cdot\bar{v}}{\lambda_{0}} t\right)}.\end{array}
\end{equation}

\subsection{Problem formulation}\label{S2.2}

Based on the system model and two channel models mentioned above, the formulation of hybrid precoding problem is discussed as follow. Our aim is to design the RIS-based hybrid precoding by maximizing the achievable sum-rate ${R}$, which can be described as
\begin{equation}\label{eq10}
{{R = }\sum\limits_{{k = }1}^{K} {{{\log }_2}\left( {1{\rm{ + }}{\gamma_{{k}}}} \right)}},
\end{equation}
where ${{\gamma _{{k}}}}$ is the signal-to-interference-plus-noise ratio (SINR) of the ${k}$th user, which is expressed as
\begin{equation}\label{eq11}
{\gamma _{k}\;=\;\frac{{{{\left| {{\bf{h}}_k^{H}{{\bf{\Phi}}}{\bf{G}}{\bf{f}}_k} \right|}^2}}}{\sum\limits_{{k^{'} \ne }k}^{K} {{\left| {{\bf{h}}_k^{H}{{\bf{\Phi}}}{\bf{G}}{\bf{f}}_{k^{'}}} \right|}^2 + {\sigma ^2}}}},
\end{equation}
where $ {\bf{f}}_k $ is the $ k $th column of digital precoding matrix ${\bf{F}} $.

As mentioned above, the aim is to maximize the achievable sum-rate $ R $ by designing the optimal the digital precoding matrix ${{\bf{F}}^{\rm{opt}}}$ and analog beamforming matrix ${{\bf{\Phi}}^{\rm{opt}}}$. The optimization problem can be formulated as
\begin{equation}\label{eq12}
{\begin{array}{l}
\left( {{\bf{\Phi}}^{\rm{opt}},{\bf{F}}^{\rm{opt}}} \right) = \mathop {\arg \max }\limits_{{{\bf{\Phi}}},\ {{\bf{F}}}} R\\
          {\rm{s.t.}} \ \ \ C_1: \left\| {\bf{F}} \right\|_{F}^2 \le \rho,\\
    \ \ \ \ \ \ \  C_2: {{\bf{\Phi}}} \in {{\bf{\cal F}}},
\end{array}}
\end{equation}
where ${{\bf{\cal F}}}$ denotes the set consisting of all possible analog beamforming matrices, and $\rho$ is the maximal transmit power. Generally, due to the practical hardware implementation, RIS can only achieve discrete phase shifts~\cite{Dai19}. We consider the 1-bit phase shift of RIS for the proposed RIS-based hybrid precoding architecture, i.e., ${ {\theta}_{m, n_s}} $ is either 0 or $ \pi $, which can be easily implemented by the diodes with very low power and cost~\cite{Dai19}. In this case, each diagonal element of ${{\bf{\Phi}}}$ in (\ref{eq8}) should be selected from ${\{ - 1, + 1\} }$.

Due to the non-convex constraint $C_2$, the optimization problem (\ref{eq12}) is non-convex. Thus, it is not a trivial task to find out the optimal solution to (\ref{eq12}). In the next section, inspired by deep learning, we will propose an optimization algorithm to achieve the near-optimal performance with low complexity.

\section{Deep Learning Based Hybrid Precoding Design}\label{S3}
In this section, the hybrid precoding optimization problem (\ref{eq12}) is reformulated first. Then we propose the DL-MDC algorithm to solve the reformulated problem. Next, we provide the complexity analysis of the DL-MDC algorithm. Finally, the insight of the idea of the DL-MDC algorithm for some other problems for wireless communications is presented. 
\subsection{Reformulation of hybrid precoding optimization problem}
The hybrid precoding optimization problem (\ref{eq12}) can be completed in two steps: digital precoding optimization and analog beamforming optimization. For the optimization of digital precoding, the low-complexity zero forcing (ZF) algorithm~\cite{Marzetta2010} can be used based on the effective channel ${{\bf{H}}_{\rm{eq}} = {\bf{H}}{{\bf{\Phi }}{\bf{G}}}}$, if the analog beamforming matrix $ \bf{\Phi } $ has been achieved, i.e., digital precoding matrix ${\bf{F}}$ can be computed as:
\begin{equation}\label{eq11}
{{\bf{F}}} = \sqrt{\rho}\frac{({\bf{H}}_{\rm{eq}})^{{H}}({\bf{H}}_{\rm{eq}}({\bf{H}}_{\rm{eq}})^{{H}})^{\rm{-1}}}{\left\| {({\bf{H}}_{\rm{eq}})^{{H}}({\bf{H}}_{\rm{eq}}({\bf{H}}_{\rm{eq}})^{{H}})^{\rm{-1}}} \right\|_{{F}}}.
\end{equation}

As we mentioned before, due to the non-convex constraint of the discrete phase shift~\cite{Dai19}, the optimization of analog beamforming is non-convex. The optimal solution to this non-convex problem is the exhaustive search algorithm, which however suffers from the unaffordable computational complexity in THz massive MIMO systems. For example, when the number of RIS elements ${N}$ is 128, the search space is ${{2^{128}} \approx 3.4 \times {10^{38}}}$, which is too large to be realized in practice. Some iterative search methods~\cite{Gao20171bit} with reduced complexity can be used to approach the optimal exhaustive search algorithm. However, the complexity of existing iterative search methods is still high for THz massive MIMO systems with numerous RIS elements. To this end, inspired by the fact deep learning can solve the classification problem with low complexity~\cite{Yuen4}, the analog beamforming optimization problem is reformulated with non-convex constraint into multiple discrete classification problems.

Specifically, as we have explained in Subsection II-B, each diagonal element of the analog beamforming matrix ${{\bm{\Phi}}} $ is either $1$ or $-1$. Thus, the mapping relationship between the channel matrix ${\textbf{H}} $ and each diagonal element in ${{\bm{\Phi}}}$ can be regarded as a binary classification problem, and accordingly the analog beamforming optimization problem can be reformulated as multiple binary classification problems, where the number of binary classification problems equals to the number of the non-zero diagonal elements of $ \bf{\Phi} $, which is $N$.

Furthermore, to figure out the binary classification pattern between the channel matrix and each diagonal element in ${{\bm{\Phi}}}$, we leverage the excellent performance of DNN for handling the classification problem to propose the DL-MDC algorithm, which can solve the reformulated sum-rate maximization problem with significantly reduced complexity and runtime.  

Specifically, the proposed DL-MDC algorithm can be divided into four steps. First, the parallel DNN system model is established. Then, we choose a traditional hybrid precoding algorithm to generate the data samples to train the parallel DNN. Finally, the trained parallel DNN system with new data samples is tested. The proposed DL-MDC algorithm with four steps will be discussed in detail in the next Subsection III-B.

\subsection{Proposed DL-MDC algorithm}
As mentioned before, the proposed DL-MDC algorithm, whose pseudo code is shown in \textbf{Algorithm 1}, is composed of four steps.

\begin{algorithm}[htbp]
\caption{Proposed DL-MDC algorithm} 
\textbf{Inputs}: The number of RIS elements: $ N $, the number of users: $K$, the number of RF chains: $ M=K $, the number of paths for ${{\bf{h}}_{k,m}}$: ${L}_{k,m}$, SNR. 
\\\hspace*{+0mm}1: \textbf{\textit{Parallel DNN system model}}: 
\\2:\hspace*{+6mm}construct a untrained parallel DNN system; 
\\3:\hspace*{+6mm}use the truncated normal distribution to initialize the neuron weights.
\\4: \hspace*{+0mm}\textbf{\textit{Data samples generation}:} 
\\5:\hspace*{+6mm}use traditional hybrid precoding algorithm to generate data samples ${\left( {{\bf{H}}^{(q)}},{{\bf{\Phi}}^{(q)}}\right) }$;
\\6:\hspace*{+0mm} \textbf{\textit{Training stage}:} 
\\7:\hspace*{+6mm}preprocess: obtain the new training tuple $ \left( \left\lbrace {{\theta_{k,n}^{(q)}}}.\right\rbrace , \left\lbrace {{\phi}^{(q)}_{n}}\right\rbrace  \right)  $;
\\8:\hspace*{+6mm}\textbf{Repeat}
\\9:\hspace*{+12mm}use the RMSprop training algorithm~\cite{RMSprop} to optimize the parallel DNN system;
\\10:\hspace*{+10mm}employ cross-validation mechanism;
\\11:\hspace*{+10mm}adjust parameters of RMSprop training algorithm;
\\12:\hspace*{+4mm}\textbf{Until} the MSEs of the validation and training sample set converge to a threshold.
\\13:\hspace*{+0mm} \textbf{\textit{Testing stage}:} 
\\14: \hspace*{+4mm}test the parallel DNN system.
\\\textbf{Output}: The trained parallel DNN system. 
\end{algorithm}

\subsubsection{\textbf{Parallel DNN system model}}
The first step of our scheme is to establish the DNN model. Through our extensive simulations, we find that the model of multiple DNNs is more stable and accurate for the binary classification problems, while the traditional single DNN may have the gradient disappearance problem due to its large number of layers and neurons. Thus, instead of using the traditional single DNN to solve the multiple binary classification problems, multiple DNNs are utilized to solve the multiple binary classification problems independently as shown in Fig. 2, where each DNN has only one output, which responds to one diagonal element of the analog beamforming matrix $ \bf{\Phi} $. In other word, each binary classification problem is solved by a particular DNN. In this way, all DNNs in the parallel DNN system can be trained and used simultaneously, so the runtime will be reduced substantially.

\begin{figure}[tb]
\begin{center}
\vspace*{0mm}\includegraphics[scale=0.45]{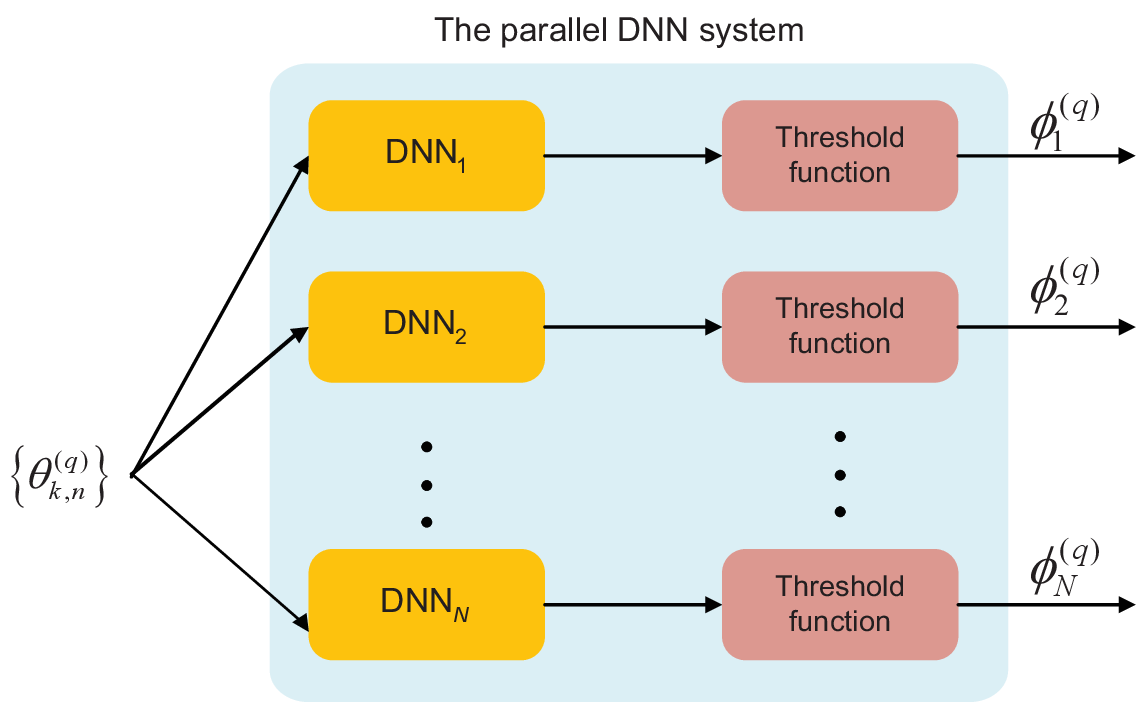}
\end{center}\vspace*{-3mm}
\vspace*{-3mm}\caption{The proposed parallel DNN system.}
\vspace*{3mm}\label{V2_3}
\end{figure}

In the proposed parallel DNN system, a nonlinear activation function named sigmoid, i.e., ${{\rm{sigmoid}}\left( x \right) = \frac{1}{{1 + {e^{ - x}}}}}$~\cite{PatternRecognition}, is adopted in the output layer. Particularly, the output of the sigmoid function ranges from 0 to 1, and the threshold is set as 0.5 to obtain the final binary classification result. 
Meanwhile, for the hidden layers of each DNN in the parallel DNN system, we choose ReLU function, i.e., ${{\mathop{\rm Re}\nolimits} {\rm{LU}}\left( x \right){\rm{ = }}\max \left( {0,x} \right)}$~\cite{PatternRecognition}, as the activation function to capture the characteristics of the data samples.

\subsubsection{\textbf{Data samples generation}}
The data samples generated by the traditional hybrid precoding algorithm are utilized as the inputs and the labels for training the parallel DNN system in Fig. 2. 
Here ``labels'' are the expected outputs of the parallel DNN system, which are used to calculate the loss function, which will be discussed in the next step of training stage. 
Specifically, the data samples are generated as follows. First, the channel matrix ${{{\bf{H}}^{(q)}}}$ is generated according to the theoretical Saleh-Valenzuela~\cite{Li18} or the practical 3GPP channel model~\cite{3GPP} as the input of the parallel DNN system, where ${q}$ denotes the index of the data sample. Secondly, based on the channel matrix ${{{\bf{H}}^{(q)}}}$, we run the traditional hybrid precoding algorithm to generate the corresponding analog beamforming matrix ${{{\bf{\Phi}}^{(q)}}}$, whose entries will be used as the labels of the parallel DNN system. As mentioned before, as 1-bit phase shift of RIS is considered for the proposed RIS-based hybrid precoding architecture in this paper, the diagonal entries in ${{{\bf{\Phi}}^{(q)}}}$ are either 1 or -1. Next, the input ${{{\bf{H}}^{(q)}}}$ and the corresponding label ${{{\bf{\Phi}}^{(q)}}}$ are combined as the training tuple ${\left( {{\bf{H}}^{(q)}},{{\bf{\Phi}}^{(q)}}\right) }$. It should be noted that we use the ZF algorithm to obtain the digital precoding matrix after the analog beamforming matrix has been achieved~\cite{DaimmWave}, where the equivalent channel used by the ZF algorithm is ${{\bf{H}}^{(q)}{\bf{\Phi }}^{(q)}{\bf{G}}^{(q)}}$.

${{\bf{\cal T}}}$ and ${{\bf{\cal V}}}$ are used to denote the indices of data samples of these two sample sets, respectively. As suggested in~\cite{WMMSE_DNN2017}, 80\% of generated data samples will be assigned to the training sample set, and 20\% to the validation sample set. These data samples will be used for training the weights of the parallel DNN system, which will be discussed in the following step.

\subsubsection{\textbf{Training stage}}
Before training the parallel DNN system, the neuron weights are initialized by the truncated normal distribution for better training performance~\cite{RMSprop}. Additionally, to mine more essential features to express the mapping relationship between the channel matrix $ { {{\bf{H}}^{(q)}} }$ and analog beamforming matrix $ {{{\bf{\Phi}}^{(q)}} }$, we also preprocess the data samples. More specifically, since the key goal of analog beamforming is to generate directional beams, the channel direction information embedded in the phase of each element in the channel matrix $ { {{\bf{H}}^{(q)}} }$, is more important than the channel gain information embedded in the amplitude of each element in the channel matrix $ { {{\bf{H}}^{(q)}} }$. Therefore, the phase $\left\lbrace {{\theta_{k,n}^{(q)}}}\right\rbrace $ of each element $\left\lbrace {{h_{k,n}^{(q)}}}\right\rbrace $ of the channel matrix ${{{\bf{H}}^{(q)}}}$ is extracted as the new input of the parallel DNN system. In addition, for the label ${{{\bf{\Phi}}^{(q)}}}$, as the sigmoid function with the output range $ \left[ 0, 1\right]  $ is adopted in the output layer, the label of the parallel DNN system, which is either 1 or -1, should be mapped to 0 or 1 to calculate the loss function~\cite{PatternRecognition}. Therefore, we use ${0.5*{{ {\rm{Diag}}\left({{{\bf{\Phi}}^{(q)}}}\right) }}+0.5}$ with the same output range $ \left[ 0, 1\right]  $, as the new label, which is denoted as ${\{{\bm{\phi}}^{(q)}_{n}\}}$, to calculate the loss function~\cite{PatternRecognition}. The updated training tuple after data samples preprocessing is denoted as $ \left( \left\lbrace {{\theta_{k,n}^{(q)}}}\right\rbrace , \left\lbrace {{\phi}^{(q)}_{n}}\right\rbrace  \right)  $.

Based on the new training tuple $ \left( \left\lbrace {{\theta_{k,n}^{(q)}}}\right\rbrace , \left\lbrace {{\phi}^{(q)}_{n}}\right\rbrace  \right)  $, we compute the typical mean square error (MSE) between the new label ${\{{\phi}^{(q)}_{n}\}}$ and the actual output of the parallel DNN system as the loss function for training, which is shown as
\begin{equation}\label{eq14}
{{{{\rm{Loss}}_{n}} =} \frac{1}{{\rm{card}}({\bf{\cal T}})}\sum_{q\in{\bf{\cal T}}}\left({{{\phi}^{(q)}_n-{y}_n}}\right)^2},
\end{equation}
where ${{{\rm{Loss}}_{n}}}$ denotes the loss function of the ${n}$th DNN, 
${{y}_n}$ denotes the actual output of the ${n}$th DNN in the parallel DNN system. 

Based on the loss function (\ref{eq14}), we use the RMSprop training algorithm~\cite{RMSprop} to optimize the weights of all neurons in the parallel DNN system. 
The decay rate of the RMSprop we choose is 0.9, as suggested in~\cite{WMMSE_DNN2017}. Meanwhile, there are some other parameters in the RMSprop algorithm, such as learning rate, batch-size, and dropout. In order to figure out the suitable parameters of the RMSprop algorithm, we employ the cross-validation mechanism in the training stage. The details of cross-validation mechanism are explained as follows. 

First, $ {\left\lbrace {{\theta_{k,n}^{(q)}}}\right\rbrace}_{q \in {\bf{\cal V}}} $ in the validation sample set are sent into the parallel DNN system at some marked epochs, which can be chosen evenly during the training stage. Then, we calculate the MSE between the label ${\{{\phi}^{(q)}_{n}\}}_{q \in {\bf{\cal V}}}$ and the actual output of the DNN. Next, based on MSEs of the validation sample set and training sample set at the marked epoch, we can adjust the parameters of the RMSprop training algorithm. For example, if the MSE of validation sample set is monotonically increasing in the training stage, it means that the parallel DNN system is over-fitted. Thus, we can increase the batch-size of the RMSprop training algorithm. Additionally, when the MSEs of validation sample set and training sample set are both too high, it means that the current parallel DNN system is not complex enough in scale to describe the considered non-convex problem. Therefore, the dropout should be reduced. Based on these new adjusted parameters of the RMSprop training algorithm, the parallel DNN system can be trained again by employing the cross-validation mechanism. The above parameter adjustment and the retraining process will stop until the MSEs of the validation and training sample sets converge to a threshold, i.e., 1\%, which means the parallel DNN system has been trained well with suitable parameters.

\subsubsection{\textbf{Testing stage}}
As mentioned in \textbf{\textit{Data samples generation}}, the parallel DNN system is trained with a large number of data samples, which contain the channel matrices across different time and space, to ensure the generalization ability. To verify the generalization capability of the parallel DNN system, we test the trained parallel DNN system by using test data samples following the same distribution as the those in the step of data samples generation. Specifically, we first pass each new channel matrix ${{{\bf{H}}}}$ through the trained parallel DNN system to get the outputs, which are the prediction results (the analog beamforming matrix) of the trained parallel DNN system. Then, the prediction results of the parallel DNN system are compared with the labels obtained from the traditional hybrid precoding algorithm to calculate the prediction accuracy of the parallel DNN system. 

\vspace{-2mm}
\subsection{Complexity analysis}
The computational complexity of proposed DL-MDC algorithm is ${{\bf{\cal O}}(N\sum\limits_{p = 0}^{P - 1} {{l_p}} {l_{p + 1}})}$, where ${{l_p}}$ denotes the number of neurons in the $p$th layer of the DNN. Meanwhile, the complexity of the optimal exhaustive searching algorithm is ${{\bf{\cal O}}(2^N)}$. Since the number of RIS elements ${N}$ is usually large, ${2^N >> N\sum\limits_{p = 0}^{P - 1} {{l_p}} {l_{p + 1}}}$ always holds. Thus, the proposed algorithm has much lower complexity than the optimal exhaustive search algorithm. 

Additionally, the complexity of the proposed DL-MDC algorithm is comparable to that of the iterative search methods for hybrid precoding. For example, the cross entropy optimization (CEO) algorithm~\cite{Gao20171bit}, which is applied as the traditional hybrid precoding algorithm with good performance in this paper, has the complexity ${{\bf{\cal O}}(ISNK^2)}$, where the number of iterations $ I $  and the number of candidates $ S $ are the parameters of the CEO algorithm. However, after the offline training of the parallel DNN system, the runtime of the parallel DNN system is much less than that of the CEO algorithm, which will be verified by simulations in Section IV.
\vspace{-2mm}
\subsection{Insights from the proposed DL-MDC algorithm}
In this paper, we reformulate the optimization problem with non-convex constraint into multiple discrete classification problems, and utilize the parallel DNN system as the efficient classifiers accordingly. This idea can be applied to solve some other problems for wireless communications. For example, the distance-frequency dependent property for THz communication~\cite{Li15}, which means the different bandwidths of transmission windows created by THz molecular absorption change drastically for different communication distances, will motivate the design of carrier assignment. The carrier assignment is to figure out the available carriers for transmission. For each carrier, whether the carrier can be used to transmit signal is a binary classification problem. Therefore, this carrier assignment design problem can be also reformulated as the multiple binary classification problems and solved by the proposed DL-MDC algorithm, where the output of each DNN in the parallel DNN system is posterior probability of the carrier being used.  

Moreover, instead of using raw data samples to train DNN directly, we preprocess the raw data samples based on the physical property of the hybrid precoding, i.e., the channel direction information is more important than the channel gain information for analog beamforming to generate directional beams. Through our extensive simulations, the prediction accuracy of the parallel DNN system trained with the processed data samples is better than that with raw data samples. This reminds us when applying deep learning techniques to solve some problems in wireless communication systems, instead of directly using the deep learning as a new tool, we can further improve the algorithm of deep learning, i.e., preprocessing the data samples or changing the activation function based on the specific problem, to achieve better performance.

\section{Simulation Results}\label{S4}
In this section, considering both the theoretical Saleh-Valenzuela channel model and the practical 3GPP channel model, the proposed DL-MDC algorithm for the proposed RIS-based hybrid precoding architecture is conducted for THz communication. The performance in terms of the prediction accuracy and runtime is evaluated via simulations. 
Moreover, the achievable sum-rate comparison between the proposed DL-MDC algorithm and the traditional hybrid precoding algorithm is presented. We choose the CEO algorithm~\cite{Gao20171bit} as the traditional hybrid precoding algorithm, which is also utilized to generate the data samples for training the parallel DNN system. Note that although the CEO algorithm is used as example, other algorithms with different performance for hybrid precoding can be approximated with less runtime by the proposed DL-MDC algorithm. 

We assume that the BS employs an RIS of $N=64~$ (or $128$)  passive elements with the element spacing ${{d_1} = {d_2} = {\lambda  \mathord{\left/{\vphantom {\lambda  2}} \right.\kern-\nulldelimiterspace} 2}}$ for the typical downlink THz massive MIMO system, and $M=4$ RF chains are used to serve $K=4$ users simultaneously. Thus, each sub-surface on the RIS consists of $N_s=16$ (or~$32$) passive elements. 

In our simulation, the proposed DL-MDC hybrid precoding scheme and the CEO algorithm are implemented in Python 3.6.0 with TensorFlow 1.3.0. Each DNN in the parallel DNN system has one input layer, three hidden layers, and one output layer.

\subsection{Simulation results in the Saleh-Valenzuela channel model}
In the simulations based on Saleh-Valenzuela channel model, 
the channel ${{{\bf{h}}_{k,m}}}$ is generated according to (2), and we assume: 1) ${{L_{k,m}} = 3}$; 2) ${{\alpha^{(l)}_{k,m}}\sim\mathcal{C}\mathcal{N}\left( {\rm{0}},\sigma^{2} \right)}$; 3) ${\varphi _{k,m}^{(l)}}$ and ${\theta _{k,m}^{(l)}}$  follow the uniform distribution ${{\mathcal{U}(-{\pi},{\pi})}}$~\cite{BM}. We consider that RIS and antenna feeder are integral parts of the transmitter, and the channel ${\bf{g}}_{m}$ between the BS and RIS is assumed to be a all-one vector of size ${{N_s}\times 1}$~\cite{transmitter3}. The first hidden layer of DNN in the parallel DNN system contains 256 neurons, and both the second and third hidden layers consist of 80 neurons.

\begin{figure}[h]
	\begin{center}
		\vspace*{-7mm}\includegraphics[scale=0.60]{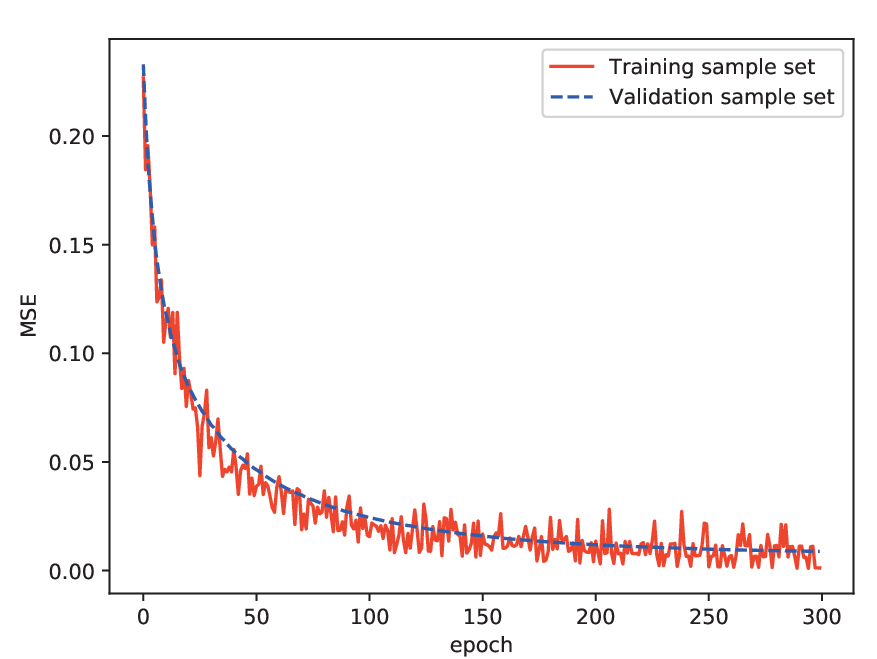}
	\end{center}
	\vspace*{-6mm}\caption{MSEs of the validation sample set and training sample set in the training stage.} \label{F7}
\end{figure}

Fig. 3 compares the MSEs of the training sample set and validation sample set. It can be observed that with the increase of training epoch, the overall trends of these two MSEs are declining. However, the MSE of validation sample set is more stable, which indicates that the generalization capability of the parallel DNN system is becoming better as the increase of the training epochs in the training stage. When the training epoch reaches about 250, the validation error converges to nearly 0.025.

\begin{table}[htbp]
	\centering
	\caption{Prediction accuracy of multiple binary classification problems $\left( N=64\right) $.}
	\label{Table1}
	\setlength{\tabcolsep}{1.58mm}
		$\left[  
		\begin{tabular}{ccccccccc}
			&  0.987 & 0.99  & 0.996 & 0.991 & 0.988 & 0.987 & 0.987 & 0.991 \\
			&  0.993 & 0.989 & 0.985 & 0.99  & 0.995 & 0.99  & 0.987 & 0.989 \\
			&  0.995 & 0.991 & 0.997 & 1     & 0.995 & 0.989 & 0.998 & 0.993 \\
			&  0.997 & 1     & 0.987 & 0.991 & 0.989 & 0.988 & 0.988 & 0.987 \\
			&  0.99  & 0.989 & 0.994 & 0.995 & 0.987 & 0.987 & 0.98  & 0.99  \\
			&  0.99  & 0.987 & 0.989 & 0.988 & 0.992 & 0.991 & 0.99  & 0.994 \\
			&  0.99  & 0.994 & 0.99  & 0.989 & 0.994 & 0.996 & 0.988 & 0.991 \\
			&  0.982 & 0.993 & 0.994 & 0.977 & 0.99  & 0.99  & 0.992 & 0.994 \\
	   \end{tabular}
	\right]$
\end{table}

In order to show the performance of the parallel DNN system in solving classification problems, we present the prediction accuracy for phase shift of each element on RIS in the testing stage in Table ${\rm\uppercase\expandafter{\romannumeral1}}$. The prediction accuracy is defined as the probability that the prediction result is the same as the CEO algorithm. 
As we have mentioned before, there are ${N}$ elements on RIS, and each element is associated with one specific DNN in the parallel DNN system. Therefore, there are ${N}$ prediction results as shown in Table ${\rm\uppercase\expandafter{\romannumeral1}}$, where the average prediction accuracy is about 0.99 for all $ N $ RIS elements. It should be noted that, the prediction accuracy of each RIS element has a slight difference due to certain randomness of mini-batch gradient descent~\cite{PatternRecognition} in the training stage.

\begin{table*}[htpb]
	\centering
	\caption{Runtime Comparison Between DL-MDC and the CEO Algorithm.}
	\begin{tabular}{|c|c|c|c|c|c|c|}
		\hline
		Number of elements on RIS          & \multicolumn{3}{c|}{$ N $=64}  & \multicolumn{3}{c|}{$ N $=128} \\ \hline
		Number of test data samples        & 100    & 500     & 1000    & 100    & 500     & 1000    \\ \hline
		Runtime of the CEO algorithm (s)   & 68.788 & 335.265 & 685.896 & 86.655 & 683.858 & 1037.01 \\ \hline
		Average runtime of one DNN (s)     & 0.015  & 0.014   & 0.024   & 0.07   & 0.063   & 0.058   \\ \hline
		Runtime of parallel DNN system (s) & 0.96   & 0.896   & 1.536   & 0.896  & 8.064   & 7.424   \\ \hline
		Time speed up                      & 72     & 374     & 447     & 10     & 85      & 120     \\ \hline
	\end{tabular}
\end{table*}

Table ${\rm\uppercase\expandafter{\romannumeral2}}$ shows the runtime comparison between the proposed DL-MDC hybrid precoding algorithm and the CEO algorithm. 
Here, the DNNs in the parallel DNN system is assumed to operate one by one. Thus, we record the runtime of only one DNN as well as the whole parallel DNN system in Table ${\rm\uppercase\expandafter{\romannumeral2}}$. If all DNNs can run at the same time, then the runtime of the parallel DNN system will reduce to that of the single one DNN. It can be seen from the Table ${\rm\uppercase\expandafter{\romannumeral2}}$ the runtime of the proposed DL-MDC hybrid precoding algorithm can be significantly reduced by about a hundred fold. For example, when $ N=128 $ and the number of test data samples is 1000, the runtime of the proposed DL-MDC algorithm is 120 times shorter than that of the CEO algorithm. 

Note that although training the parallel DNN system with data samples requires a lot of time, since a large number of channel matrices across different time and space should be used to improve the generalization capability, the training stage is realized offline. Therefore, once the parallel DNN system has been trained well, the runtime of the trained parallel DNN system is much less than that of the CEO algorithm. 

\vspace*{0mm}
\begin{figure*}[hbp] 
	\centering
	\subfigure[]{
		\begin{minipage}[b]{0.8\textwidth}
			~~~~~~~\includegraphics[width=0.8\textwidth]{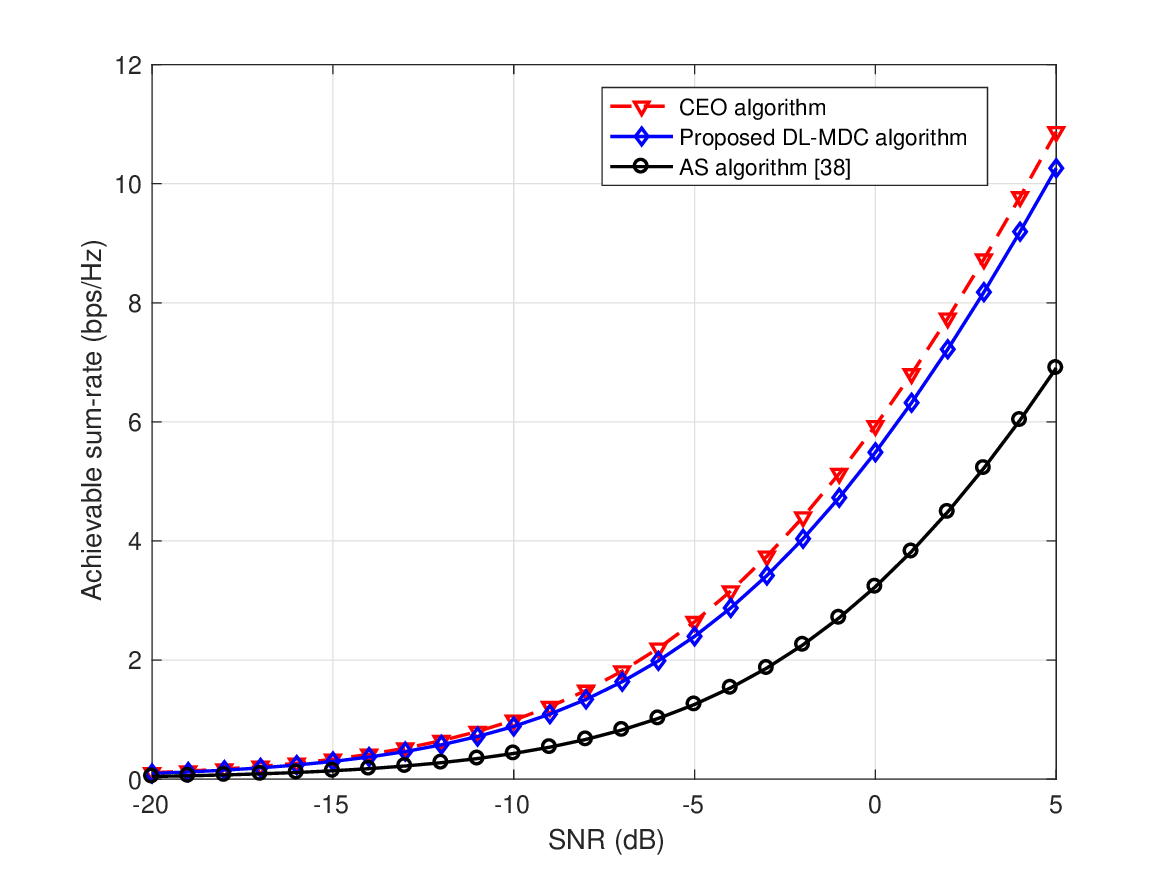}
		\end{minipage}
	}

	\subfigure[]{
		\begin{minipage}[b]{0.8\textwidth}
			~~~~~~~\includegraphics[width=0.8\textwidth]{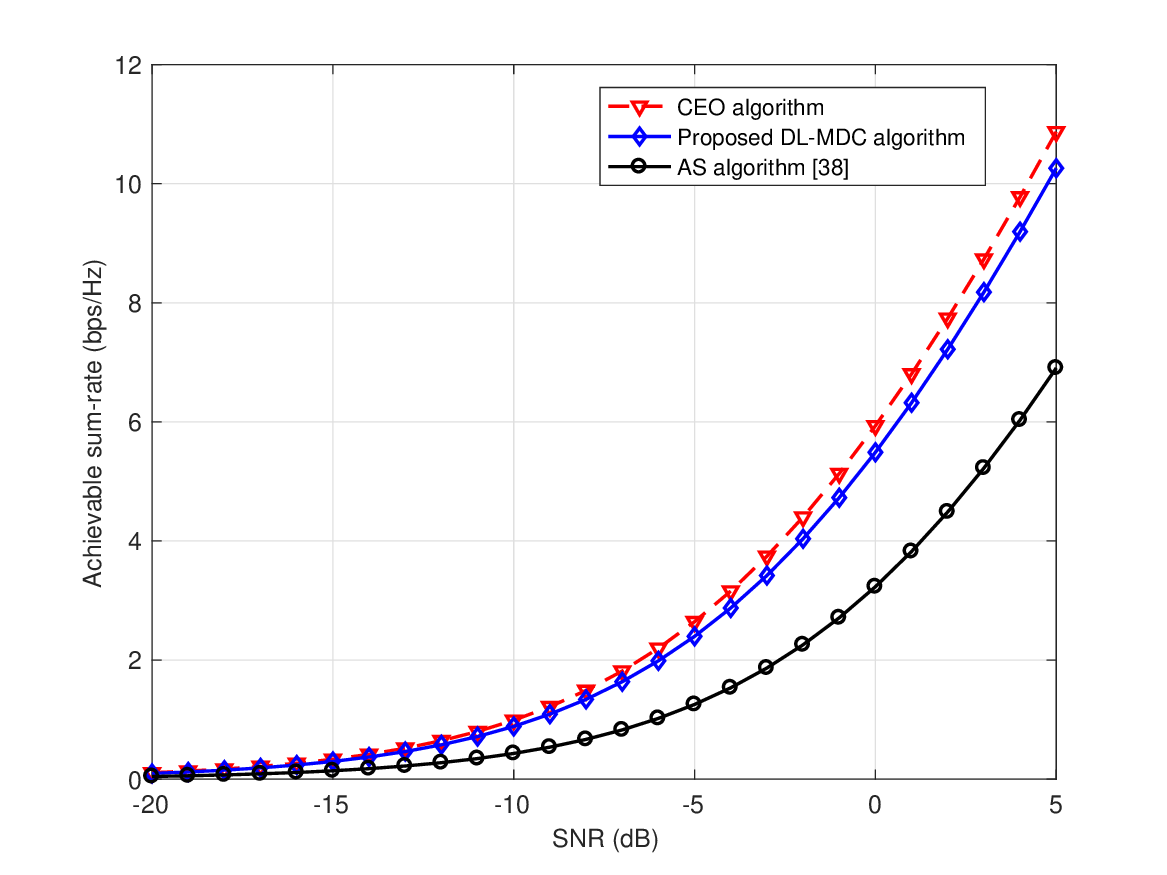}
		\end{minipage}
	}
	\caption{Achievable sum-rate comparison: (a) ${N}$=64; (b) ${N}$=128.}
\end{figure*}
 
Fig. 4 presents the comparison of the achievable sum-rate performance in a typical THz massive MIMO system with  ${K = M = 4}$, where ${N = 64}$ for Fig. 4 (a) and ${N = 128}$ for Fig. 4 (b). The following three schemes are compared in Fig. 4. The first scheme is the proposed DL-MDC algorithm, which is designed for RIS-based THz massive MIMO system. The second scheme is the CEO algorithm, which is the traditional hybrid precoding algorithm to generate data samples. The proposed DL-MDC algorithm tries to approximate the CEO algorithm by using deep learning with much lower complexity. Thus, the performance of the CEO algorithm is the upper bound of the proposed DL-MDC algorithm. 
Finally, the conventional antenna selection (AS)-based hybrid precoding scheme~\cite{power} is the third scheme, which can be adapted to the proposed RIS-based hybrid precoding architecture for performance comparison. It can be observed from Fig. 4 that the proposed DL-MDC algorithm outperforms the traditional AS-based hybrid precoding scheme. Additionally, the performance of the proposed DL-MDC hybrid precoding algorithm is close to the CEO algorithm. For example, at SNR = 5 dB in Fig. 4 (a), the achievable  sum-rate of the proposed DL-MDL algorithm is about 95\% of that of the CEO algorithm, while the runtime of the proposed DL-MDL algorithm can be significantly reduced as shown in Table ${\rm\uppercase\expandafter{\romannumeral2}}$.  

\begin{figure}[h]
	\begin{center}
		\vspace*{0mm}\includegraphics[width=0.65\linewidth]{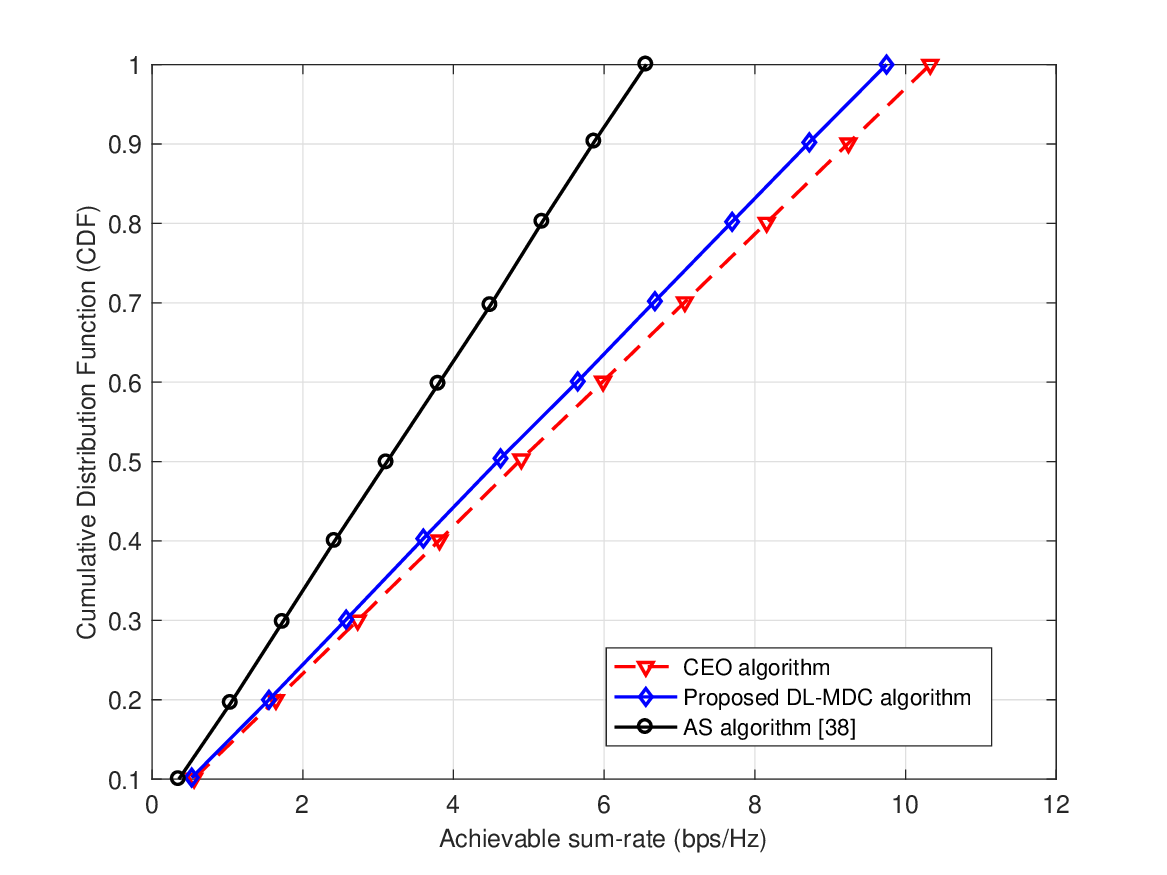}
	\end{center}
	\vspace*{-3mm}\caption{Comparison of the cumulative distribution function (CDF) of the achievable sum-rate.} \label{F6}
\end{figure}

Fig. 5 shows the cumulative distribution function (CDF) of the achievable  sum-rate of the three schemes mentioned above, where ${N=64}$, ${M=K=4}$. It is observed that the achievable sum-rate performance of the proposed DL-MDC algorithm is very close to that of the CEO algorithm. Meanwhile, both schemes significantly outperform from the AS-based hybrid precoding scheme.

\subsection{Simulation results in the 3GPP channel model}

It is clear from Subsection II-B that the 3GPP channel model of rich scattering environment is much more complicated than the typical theoretical Saleh-Valenzuela channel model. Therefore, for the parallel DNN system model with the same scale as mentioned in Subsection IV-A, mining the mapping relationship between the channel matrix $ { {{\bf{H}}^{(q)}} }$ generated by the 3GPP channel model and analog beamforming matrix $ {{{\bf{\Phi}}^{(q)}} }$ is more difficult than that generated by the Saleh-Valenzuela channel model. In this case, the scale of the parallel DNN system based on the Saleh-Valenzuela channel model in Subsection IV-A needs to be expanded to approximate the mapping function of the channel matrix $ { {{\bf{H}}^{(q)}} }$ generated by the 3GPP channel model and analog beamforming matrix $ {{{\bf{\Phi}}^{(q)}} }$. Here, in the simulation based on the 3GPP channel model, the three hidden layers contain 512, 256, 128 neurons, respectively.

\begin{figure}[h]
	\begin{center}
		\vspace*{-7mm}\includegraphics[scale=0.65]{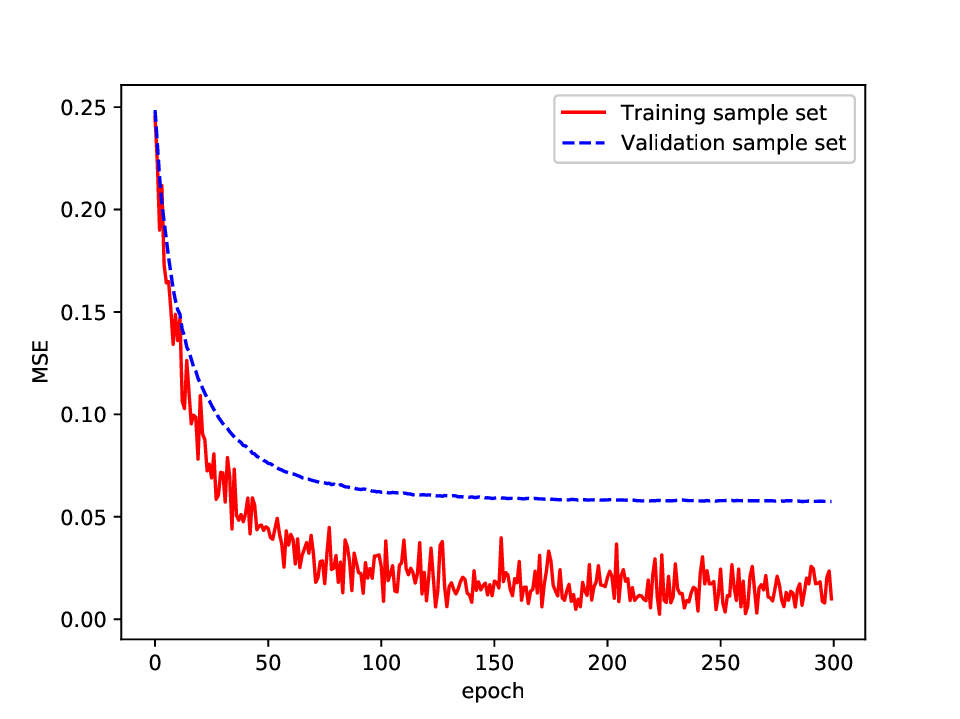}
	\end{center}
	\vspace*{-6mm}\caption{MSEs of the validation sample set and training sample set in the training stage.} \label{F7}
\end{figure}

\begin{figure}[h]
	\begin{center}
		\vspace*{0mm}\includegraphics[width=0.6\linewidth]{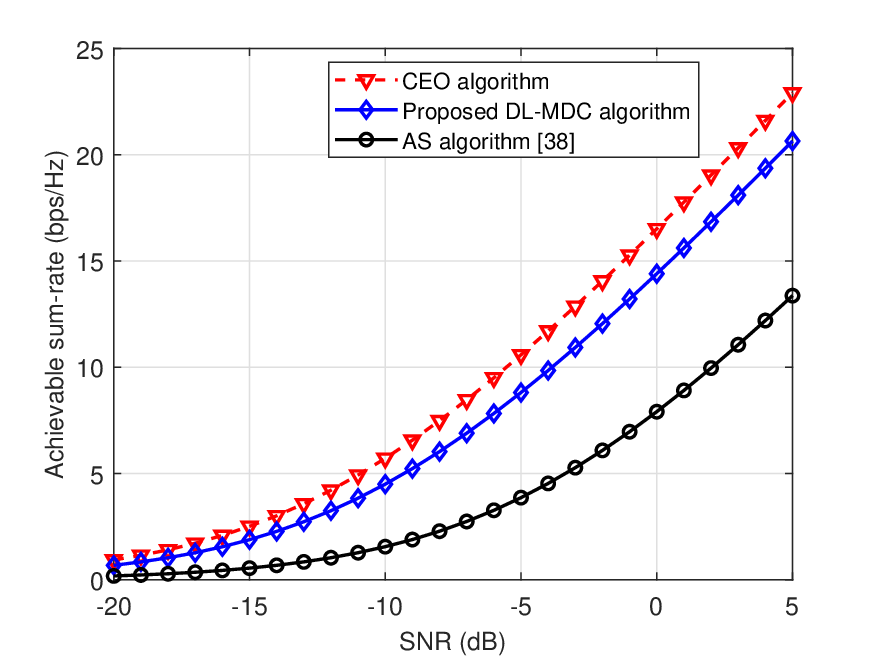}
	\end{center}
	\vspace*{-3mm}\caption{Achievable sum-rate comparison.}
	\vspace{3mm} 
\end{figure}
\vspace{2mm}
Fig. 6 compares the MSEs of the training sample set and validation sample set. Similar to Fig.3, these two MSEs are also declining as the increase of training epoch. Meanwhile, the MSE of validation sample set converges to nearly 0.05, which is a little bit larger than that in Fig. 3. Thus, the average prediction accuracy 0.94 is lower than the average prediction accuracy 0.99 in Fig. 3. The influence of this small difference in the average prediction accuracy between Saleh-Valenzuela channel model and 3GPP channel model will be considered and discussed in simulation results in the following Fig.7.

Fig. 7 shows the achievable sum-rate performance comparison in a THz massive MIMO system with parameters ${K = M = 4}$ and ${N = 64}$, which are same as those for the Saleh-Valenzuela channel model in Subsection IV-A. The same three schemes considered in Subsection IV-A are compared. Similarly, it can be seen that the proposed DL-MDC algorithm outperforms the conventional AS-based hybrid precoding scheme. Meanwhile, and the proposed DL-MDC algorithm can also approach to the CEO algorithm. For example, at SNR = 5 dB, the achievable sum-rate of the proposed DL-MDL algorithm is about 91\% of that of the CEO algorithm. Therefore, the small difference in the average prediction accuracy shown in Fig. 6, will not affect the sum-rate performance of the proposed DL-MDC algorithm too much. 

\section{Conclusions}\label{S5}
In this paper, we proposed a RIS-based hybrid precoding architecture for THz communication, where the energy-efficient RIS instead of the energy-hungry phased array was used to realize the analog beamforming of the hybrid precoding. In order to maximize the sum-rate of the users, the hybrid precoding matrix was carefully designed. Specifically, we reformulated the sum-rate optimization problem as multiple discrete classification problems because of the non-convex constraint of discrete phase shifts in RIS-based hybrid precoding architecture. Then, the DL-MDC hybrid precoding algorithm was proposed to solve the multiple discrete classification problems. The simulation result showed that the proposed DL-MDC hybrid precoding algorithm has much lower runtime than the traditional hybrid precoding algorithm with negligible the achievable sum-rate performance. Thus, the parallel DNN system utilized in the proposed DL-MDC algorithm can approximate the complex non-convex function of the traditional hybrid precoding algorithm well, which provides a reasonable trade-off between performance and complexity. 

\vspace{5mm}
\bibliography{IEEEabrv,LuRef}
\end{document}